\documentclass[aps,prc,amsmath,two column,amssymb,showpacs]{revtex4}
\usepackage{txfonts}
\usepackage{mathbbold}
\usepackage{epsfig,bm,dcolumn}
\usepackage{graphicx}
\usepackage{color}

\begin{document}

\title{Nonperturbative effects on the ferromagnetic transition in repulsive Fermi gases}

\author{Lianyi He$^{1,2,}$}
\email{lianyi@itp.uni-frankfurt.de}

\author{Xu-Guang Huang$^{1,}$}
\email{xhuang@itp.uni-frankfurt.de}

\affiliation{1 Institut f{\"u}r Theoretische Physik, Goethe-Universit{\"a}t, 60438 Frankfurt am
Main, Germany\\
2 Frankfurt Institute for Advanced Studies, Goethe-Universit{\"a}t, 60438 Frankfurt am
Main, Germany}

\date{\today}

\begin{abstract}
It is generally believed that a dilute spin-$\frac{1}{2}$ Fermi gas
with repulsive interactions can undergo a ferromagnetic phase
transition to a spin-polarized state at a critical gas parameter
$(k_{\rm F}a)_c$. Previous theoretical predictions of the
ferromagnetic phase transition have been based on the perturbation
theory, which treats the gas parameter as a small number. On the
other hand, Belitz, Kirkpatrick, and Vojta (BKV) have argued that
the phase transition in clean itinerant ferromagnets is generically
of first order at low temperatures, due to the correlation effects
that lead to a nonanalytic term in the free energy. The second-order
perturbation theory predicts a first-order phase transition at
$(k_{\rm F}a)_c=1.054$, consistent with the BKV argument. However,
since the critical gas parameter is expected to be of order $O(1)$,
perturbative predictions may be unreliable. In this paper we study
the nonperturbative effects on the ferromagnetic phase transition by
summing the particle-particle ladder diagrams to all orders in the
gas parameter. We consider a universal repulsive Fermi gas where the
effective range effects can be neglected, which can be realized in a
two-component Fermi gas of $^6$Li atoms by using a nonadiabatic
field switch to the upper branch of a Feshbach resonance with a
positive $s$-wave scattering length. Our theory predicts a
second-order phase transition, which indicates that ferromagnetic
transition in dilute Fermi gases is possibly a counterexample to the
BKV argument. The predicted critical gas parameter $(k_{\rm
F}a)_c=0.858$ is in good agreement with the recent quantum Monte
Carlo result $(k_{\rm F}a)_c=0.86$ for a nearly zero-range potential
[S. Pilati, \emph{et al}., Phys. Rev. Lett. {\bf 105}, 030405
(2010)]. We also compare the spin susceptibility with the quantum
Monte Carlo result and find good agreement.
\end{abstract}


\maketitle

\section{Introduction}

Itinerant ferromagnetism  is a fundamental problem in
condensed-matter physics, which can be dated back to the basic model
proposed by Stoner \cite{Stoner}. While the problem of itinerant
ferromagnetism in electronic systems is quite complicated and the
phase transition theory is still qualitative, a dilute
spin-$\frac{1}{2}$ Fermi gas with repulsive interactions may serve
as a clean system to simulate the Stoner model. It is generally
thought that the repulsive Fermi gas could undergo a ferromagnetic
phase transition (FMPT) to a spin-polarized state with increased
interaction strength \cite{Huang}. Recently, the experimentalists
realized a two-component ``repulsive" Fermi gas of $^6$Li atoms in a
harmonic trap by using a nonadiabatic field switch to the upper
branch of a Feshbach resonance with a positive $s$-wave scattering
length \cite{exp,exp2}. Therefore, it is possible to investigate
itinerant ferromagnetism in cold Fermi gases. The experimental
progress in this direction has attracted intense theoretical
interest
\cite{QMC,QMC2,FMPT1,FMPT1-2,FMPT1-3,FMPT1-4,FMPT1-5,FMPT1-6,FMPT1-7,FMPT2,FMPT3,FMPT4,FMPT5,FMPT6,FMPT7,FMPT8,
FMPT9,FMPT10,FMPT11,FMPT12,FMPT13,FMPT14,FMPT15,FMPT16,FMPT17,FMPT18,FMPT19,FMPT20,FMPT21,FMPT22,FMPT23,FMPT24}.

The physical picture of the ferromagnetism in
repulsive Fermi gases can be understood as a result of the
competition between the repulsive interaction and the Pauli
exclusion principle. The former tends to induce polarization and
reduce the interaction energy, while the latter prefers
balanced spin populations and hence a reduced kinetic energy. With increasing repulsion,
the reduced interaction energy for a polarized state will
overcome the gain in kinetic energy, and a FMPT should occur when the minimum of the energy
landscape shifts to nonzero polarization or magnetization.

Quantitatively, to study the FMPT in dilute Fermi gases at zero
temperature, we should calculate the energy density $\cal{E}$ as a
function of the spin polarization or magnetization
$x=(n_\uparrow-n_\downarrow)/(n_\uparrow+n_\downarrow)$ at given
dimensionless gas parameter $k_{\text F}a$ which represents the
interaction strength \cite{Huang}. Here, $k_{\text F}$ is the Fermi
momentum related to the total density $n=n_\uparrow+n_\downarrow$ by
$n=k_{\text F}^3/(3\pi^2)$ and $a>0$ is the $s$-wave scattering
length. Generally, the energy density can be expressed as
\begin{eqnarray}
{\cal E}(x)=\frac{3}{5}nE_{\text F}f(x),
\end{eqnarray}
where $E_{\text F}=k_{\text F}^2/(2M)$ is the Fermi energy with $M$
being the fermion mass. The dimensionless function $f(x)$, which
depends on the gas parameter $k_{\text F}a$, represents the energy
landscape with respect to the magnetization $x$.

For the order of the FMPT, Belitz, Kirkpatrick, and Vojta (BKV)
\cite{Belitz} have argued that the phase transition in clean
itinerant ferromagnets is generically of first order at low
temperatures, due to the correlation effects or the coupling of the
order parameter to gapless modes that lead to a nonanalytic term in
the free energy. The general form of the Ginzburg-Landau free energy
for clean itinerant ferromagnets takes the form
\begin{eqnarray}
f_{\text{GL}}(x)=\alpha x^2+\upsilon x^4\text{ln}|x|+\beta
x^4+O(x^6),
\end{eqnarray}
where we can keep $\beta>0$. If the coefficient $\upsilon$ is
positive, the phase transition is always of first order. On the
other hand, for negative $\upsilon$,  one always has a second-order
phase transition. The BKV argument is based on the assumption
$\upsilon>0$, motivated by perturbation theory \cite{Belitz}. This
is true for many solid-state systems where the FMPT occurs at weak
coupling. However, for dilute Fermi gases where the critical gas
parameter is expected to be of order $O(1)$, the assumption of a
positive $\upsilon$ is not reliable.

In this paper, we study the nonperturbative effects on the FMPT by
summing a set of particle-particle ladder diagrams to all orders in
the gas parameter, motivated by the large-dimension expansion
proposed by Steele \cite{resum1}. We consider a universal repulsive
Fermi gas where the effective range effect can be neglected,
corresponding to a two-component ``upper branch" Fermi gas with a
positive $s$-wave scattering length. The prediction may also apply
to the hard-sphere gas since the effective range corrections are
sub-leading-order contributions in the large-dimension expansion.
Our main conclusions for the order and the critical gas parameter of
the FMPT can be summarized as follows:
\\ (1) \emph{Order of phase transition}. We predict a second-order
phase transition, in contrast to the BKV argument. This suggests
that the FMPT in dilute Fermi gas may correspond to the case of
negative $\upsilon$.
\\ (2) \emph{Critical gas parameter}. We predict a critical gas
parameter $(k_{\text F}a)_c=0.858$ where the spin susceptibility
$\chi$ diverges. The critical gas parameter and the spin
susceptibility we obtained are in good agreement with the quantum
Monte Carlo results \cite{QMC}.

The paper is organized as follows. In Sec. II we briefly review the
perturbative predictions for FMPT in dilute Fermi gases. In Sec. III
we introduce the effective field theory approach to the two-body
scattering problem and show how we can recover the scattering
amplitude by ladder resummation. We study the nonpertuabtive effects
on FMPT in the theory of ladder resummation in Sec. IV and
investigate the role of hole-hole ladders in Sec. V. We summarize in
Sec. VI.

\section{Perturbative Predictions}

In the perturbation theory, the gas parameter $k_{\text F}a$ is
treated as a small number. Up to the order $O((k_{\text F}a)^2)$,
the expression for $f(x)$ is universal, that is, independent of the
details of the short-range interaction. We have
\begin{eqnarray}
f(x)=\frac{1}{2}(\eta_\uparrow^5+\eta_\downarrow^5)+\frac{10k_{\text F}a}{9\pi}\eta_\uparrow^3
\eta_\downarrow^3+\frac{(k_{\text F}a)^2}{21\pi^2}\xi(\eta_\uparrow,\eta_\downarrow),\label{second}
\end{eqnarray}
where $\eta_\uparrow=(1+x)^{1/3}$ and $\eta_\downarrow=(1-x)^{1/3}$.
The zeroth-order term corresponds to the kinetic energy, and the
first-order term coincides with the Hartree-Fock mean-field theory
\cite{Huang}. The coefficient $\xi(\eta_\uparrow,\eta_\downarrow)$
in the second-order term was first evaluated by Kanno \cite{Kanno}.
Its explicit form is
\begin{eqnarray}
\xi&=&22\eta_\uparrow^3\eta_\downarrow^3(\eta_\uparrow+\eta_\downarrow)-4\eta_\uparrow^7\text{ln}\frac{\eta_\uparrow+\eta_\downarrow}{\eta_\uparrow}-4\eta_\downarrow^7\text{ln}\frac{\eta_\uparrow+\eta_\downarrow}{\eta_\downarrow}\nonumber\\
&+&\frac{1}{2}(\eta_\uparrow-\eta_\downarrow)^2\eta_\uparrow\eta_\downarrow(\eta_\uparrow+\eta_\downarrow)[15(\eta_\uparrow^2+\eta_\downarrow^2)+11\eta_\uparrow\eta_\downarrow]\nonumber\\
&+&\frac{7}{4}(\eta_\uparrow-\eta_\downarrow)^4(\eta_\uparrow+\eta_\downarrow)(\eta_\uparrow^2+\eta_\downarrow^2+3\eta_\uparrow\eta_\downarrow)\text{ln}\bigg|\frac{\eta_\uparrow-\eta_\downarrow}{\eta_\uparrow+\eta_\downarrow}\bigg|.
\end{eqnarray}
Setting $x=0$, we recover the well-known equation of state for
dilute Fermi gases,
\begin{eqnarray}
{\cal E}=\frac{3}{5}nE_{\text F}\left[1+\frac{10}{9\pi}k_{\text
F}a+\frac{4(11-2\ln2)}{21\pi^2}(k_{\text F}a)^2\right],
\end{eqnarray}
which was first obtained by Huang, Yang, and Lee \cite{HYL} and
recovered by Hammer and Furnstahl \cite{EFT} in recent years using
effective field theory.

In the first-order perturbation, the FMPT is of second order and
occurs at $k_{\text F}a=\pi/2$ \cite{Huang}. However, taking into
account the second-order corrections, one finds a first-order FMPT
at $k_{\text F}a=1.054$ \cite{FMPT2,FMPT3}. This can be understood
by noticing the nonanalytical term $\propto x^4\text{ln}|x|$ with
positive coefficient in the small-$x$ expansion of the coefficient
$\xi(\eta_\uparrow,\eta_\downarrow)$. The small-$x$ expansion of the
energy density (3) takes the form
\begin{eqnarray}
f(x)=f(0)+\alpha x^2+\upsilon x^4\text{ln}|x|+\beta x^4+O(x^6),
\end{eqnarray}
which is consistent with the BKV argument \cite{Belitz} that the
correlation effects or the coupling of the order parameter to
gapless modes generally leads to nonanalytical terms in the free
energy. The coefficient $\upsilon$ can be evaluated as
\begin{eqnarray}
\upsilon=\frac{40(k_{\text F}a)^2}{243\pi^2}.
\end{eqnarray}
Therefore, up to the order $O((k_{\text F}a)^2)$, the Fermi gas
problem corresponds to the case $\upsilon>0$ which is assumed in the
BKV argument.

In general, we expect that the critical parameter is of order
$O(1)$. Therefore, the perturbative predictions for the FMPT are
probably unreliable. There naturally arises a serious problem: Does
the dilute Fermi gas problem really correspond to the case
$\upsilon>0$ if the nonperturbative effects at $k_{\text F}a\sim
O(1)$ are taken into account? For the two-body problem in the
vacuum, it is well known that an infinite set of bubble diagrams
with the leading-order contact interaction must be resummed in order
to reproduce the correct scattering amplitude if the two-body
scattering length is large \cite{Kaplan}. Therefore it is natural to
extend the idea of ladder resummation to finite density so that the
predicted equation of state works well even at $k_{\text F}a\sim
O(1)$. We can also compare the non-perturbative predictions with the
results from recent quantum Monte Carlo (QMC) simulations
\cite{QMC,QMC2}.

\section{Ladder Resummation for Two-Body Scattering }
Before we establish a nonperturbative description for the FMPT in
dilute Fermi gases, it is instructive to start with the low-energy
effective field theory in vacuum \cite{EFT, Kaplan} and to see how
the two-body scattering amplitude is reproduced from the ladder
resummation method.

For nonrelativistic two-body scattering in the $s$-wave channel
associated with a short-range interaction, the scattering amplitude
${\cal A}(k)$ is related to the $s$-wave scattering phase shift
$\delta$ by
\begin{eqnarray}
{\cal A}(k)=-\frac{4\pi}{M}\frac{1}{k\cot\delta-ik},
\end{eqnarray}
where $k\equiv|{\bf k}|$ is the scattering momentum in the
center-of-mass frame. If there exist bound states for attractive
interactions, the scattering amplitude should exhibit some imaginary
poles, $k=i\sqrt{-ME_{\text b}}$, on the complex $k$ plane with
$E_{\text b}<0$ being the binding energy. In general, the
short-range interaction is characterized by a momentum scale
$\Lambda$. Therefore, for low-energy scattering, that is, $k\ll
\Lambda$, the quantity $k\cot\delta$ can be expanded as a Taylor
series in $k^2/\Lambda^2$. In quantum scattering theory, this is
called the effective range expansion,
\begin{eqnarray}
k\cot\delta&=&-\frac{1}{a}+\frac{1}{2}\sum_{n=0}^\infty
r_n\Lambda^2\left(\frac{k^2}{\Lambda^2}\right)^{n+1}\nonumber\\
&=&-\frac{1}{a}+\frac{1}{2}r_0k^2+\ldots,
\end{eqnarray}
where $a$ is the scattering length and $r_0$ is the effective range.
For a natural system \cite{Kaplan}, we have $|a|\sim 1/\Lambda$ and
$|r_n|\sim1/\Lambda$. An example commonly studied is a hard-sphere
gas with radius $R$, in which case $a=R$ and $r_0=2R/3$. For cold
atomic gases the interatomic interaction can be tuned by means of
the Feshbach resonance, and we can have $|a|\gg |r_n|\sim
1/\Lambda$.

According to the effective range expansion, one can construct the
low-energy effective field theory \cite{EFT, Kaplan} describing
scattering at momenta $k\ll \Lambda$. Since we assume $k\ll
\Lambda$, all interactions in the effective Lagrangian are contact
interactions. The low-energy effective Lagrangian contains infinite
contact interaction terms and is given by \cite{EFT}
\begin{eqnarray}
{\cal L}_{\text{eff}}
&=&\psi^{\dagger}\left(i\partial_t+\frac{\overrightarrow{\nabla}^2}{2M}\right)
\psi-\frac{C_0}{2}(\psi^{\dagger}\psi)^2\nonumber\\
&+&\frac{C_2}{16}\left[\left(\psi\psi\right)^\dagger(\psi\tensor{\nabla}^2\psi)+\rm{H.c.}\right]+\ldots,
\end{eqnarray}
where $C_0$ and $C_2$ are dimensionful coupling constants,
$\tensor{\nabla}=\overrightarrow{\nabla}-\overleftarrow{\nabla}$ is
a Galilei invariant derivative, and $\ldots$ denotes interactions
with more derivatives ($\sim \nabla^{2n}$, $n\geq 2$) which
generally have coupling constants $C_{2n}$. The coupling constants
$C_{2n}$ ($n=0,1,2,...$) should be determined by reproducing the
scattering amplitude ${\cal A}(k)$.

In practice, we can reproduce the scattering amplitude ${\cal A}(k)$
order by order in a Taylor expansion in $k/\Lambda$. For small
scattering length ($|a|\sim 1/\Lambda$ and $|ak|\ll 1$), we can
expand the scattering amplitude as
\begin{eqnarray}
{\cal A}(k)=\frac{4\pi
a}{M}\left[1-iak+\left(\frac{ar_0}{2}-a^2\right)k^2+\ldots\right].
\end{eqnarray}
However, for large scattering length ($|a|\gg 1/\Lambda$), Kaplan
\emph{et al.} showed that one needs to expand ${\cal A}(k)$ in
powers of $k/\Lambda$ while retaining $ak$ to all orders
\cite{Kaplan}:
\begin{eqnarray}
\label{expansionofA} {\cal
A}(k)=\frac{4\pi}{M}\frac{1}{1/a+ik}\left[1+\frac{r_0/2}{1/a+ik}k^2+\ldots\right].
\end{eqnarray}
This means if the scattering length is large, the loop diagrams with
the leading-order interaction $C_0$ have to be resummed.

\begin{figure}[!htb]
\begin{center}
\includegraphics[width=9.3cm]{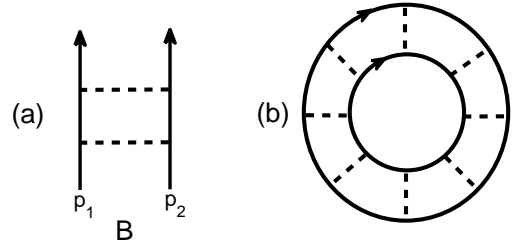}
\caption{(a) The elementary particle-particle bubble $B(p_1,p_2)$
with external momenta $p_1$ and $p_2$ for the two spin components.
The solid line with arrow corresponds to the particle term of the
propagator (\ref{propagator}). The dashed line represents the
interaction vertex $C_0$. (b) A typical particle-particle ladder
diagram contributing to the interaction energy. \label{fig1}}
\end{center}
\end{figure}

According to the free fermion propagator ${\cal G}_0(p_0,{\bf
p})=1/(p_0-\omega_{\bf p}+i\epsilon)$ with the free dispersion
$\omega_{\bf p}={\bf p}^2/(2M)$, the one-loop bubble diagram
$B_0(P_0,{\bf P})$ [Fig.\ref{fig1}a] is given by
\begin{widetext}
\begin{eqnarray}
\label{vacuumbubble}
B_0(P_0,{\bf
P})=i\int\frac{d^4q}{(2\pi)^4}\frac{1}{\frac{P_0}{2}+q_0-\frac{({\bf
P}/2+{\bf
q})^2}{2M}+i\epsilon}\frac{1}{\frac{P_0}{2}-q_0-\frac{({\bf
P}/2-{\bf q})^2}{2M}+i\epsilon}=\int\frac{d^3{\bf
q}}{(2\pi)^3}\frac{1}{P_0-\frac{{\bf P}^2}{4M}-\frac{{\bf
q}^2}{M}+i\epsilon}.
\end{eqnarray}
\end{widetext}
Here $P_0$ and ${\bf P}$ are the total energy and momentum of the
pairs in the bubble diagram [see Fig.\ref{fig1}(a)]. Let ${\bf p}_1$
and ${\bf p}_2$ be the momenta of the scattering fermions, and we
have ${\bf P}={\bf p}_1+{\bf p}_2$ and ${\bf k}=({\bf p}_1-{\bf
p}_2)/2$. Further, if the on-shell condition $P_0=({\bf p}_1^2+{\bf
p}_2^2)/(2M)={\bf P}^2/(4M)+{\bf k}^2/M$ is imposed, we find that
$B_0$ depends only on the relative momentum ${\bf k}$, corresponding
to the translational invariance.

The integral over ${\bf q}$ in Eq. (\ref{vacuumbubble}) is linearly
divergent and therefore needs to be regularized. A natural
regularization scheme is to use a momentum cutoff equal to $\Lambda$
\cite{Hammer}. In this paper, we employ the dimensional
regularization scheme. To this end, we change the space-time
dimension from $4$ to $D$ and multiply the integral by a factor
$(\mu/2)^{D}$. Here $\mu$ is an arbitrary mass scale introduced to
allow the couplings $C_{2n}$ multiplying operators containing
$\nabla^{2n}$ to have same dimensions for any $D$.  In general, the
integral $B_0$ in $D$ dimension can be evaluated as \cite{Kaplan}
\begin{eqnarray}
B_0(P_0,{\bf
P})&=&-\Gamma\left(\frac{3-D}{2}\right)\frac{(\mu/2)^{4-D}}{(4\pi)^{(D-1)/2}}\nonumber\\
&\times&M\left(-MP_0+\frac{{\bf P}^2}{4}-i\epsilon\right)^{(D-3)/2}.
\end{eqnarray}
For small scattering length, it is convenient to use the minimal
subtraction (MS) scheme which subtracts any $1/(D-4)$ pole before
taking the $D\rightarrow4$ limit. However, for large scattering
length, it is more convenient to use the power divergence
subtraction (PDS) scheme. The PDS scheme involves subtracting from
the dimensionally regularized loop integrals not only the $1/(D-4)$
poles corresponding to log divergences, as in MS, but also poles in
lower dimensions which correspond to power law divergences at $D=4$.
The integral $B_0$ has a pole in $D=3$ dimensions. It can be removed
by adding a counterterm $\delta B_0=M\mu/[4\pi(3-D)]$ to $B_0$
\cite{Kaplan}. Finally, the subtracted integral in $D=4$ dimensions
is
\begin{eqnarray}
B_0(P_0,{\bf P})=-\frac{M}{4\pi}\left(\mu-\sqrt{-MP_0+\frac{{\bf
P}^2}{4}-i\epsilon}\right).
\end{eqnarray}
Note that the MS scheme corresponds to the $\mu=0$ case.

The dependence of $C_{2n}(\mu)$ on $\mu$ is determined by the
requirement that the scattering amplitude is independent of the
arbitrary mass scale $\mu$. To this end, we impose the on-shell
condition, $P_0={\bf P}^2/(4M)+{\bf k}^2/M$. Then the one-loop
bubble diagram becomes $B_0(k)=-M(\mu+ik)/(4\pi)$. Summing the
bubble diagrams with $C_0$ vertices, we obtain \cite{Kaplan}
\begin{eqnarray}
{\cal
A}(k)=\frac{C_0(\mu)}{1-C_0(\mu)B_0(k)}+\frac{C_2(\mu)k^2}{\left[1-C_0(\mu)B_0(k)\right]^2}+\ldots.
\end{eqnarray}
Comparing this result with the expansion (\ref{expansionofA}), we
obtain
\begin{eqnarray}
C_0(\mu)&=&\frac{4\pi}{M}\frac{1}{-\mu+1/a},\nonumber\\
C_2(\mu)&=&\frac{4\pi}{M}\left(\frac{1}{-\mu+1/a}\right)^2\frac{r_0}{2}.
\end{eqnarray}
It was shown that these results fulfill the renormalization group
equations \cite{Kaplan}. We note that the mass scale $\mu$ is
similar to the cutoff $\Lambda$. In the cutoff scheme, we have
$C_0(\Lambda)=(4\pi/M)(-2\Lambda/\pi+1/a)$ \cite{Hammer}.

In the following, we mainly consider a short-range potential with a
positive scattering length $a$ and negligible effective range
$r_0\ll a$. In this case, we are able to obtain a universal result
for $f(x)$ which is independent of the details of the interaction.
In this case, the pair propagator ${\cal S}_0(P_0,{\bf P})$ in the
vacuum is given by
\begin{eqnarray}
{\cal S}_0(P_0,{\bf P})&=&\frac{C_0(\mu)}{1-C_0(\mu)B_0(P_0,{\bf
P})}\nonumber\\
&=&\frac{4\pi}{M}\frac{1}{1/a-\sqrt{-MP_0+\frac{{\bf
P}^2}{4}-i\epsilon}}.
\end{eqnarray}
For positive scattering length, the pair propagator has a pole given
by $P_0=-1/(Ma^2)+{\bf P}^2/(4M)$. This pole corresponds to a bound
state with binding energy $E_{\text b}=-1/(Ma^2)$ and effective mass
$2M$. Therefore, if the effective range is negligible, the
underlying potential must be attractive and the ground state is a
bound molecule of size $a$. However, for two-body scattering state
with positive center-of-mass energy $E={\bf k}^2/M>0$, the effective
force is repulsive. This is the so-called ``upper branch," which is
well defined in the two-body picture. For the many-body problem, a
metastable ``repulsive" Fermi gas can be realized if all fermions
are forced on the upper branch of a Feshbach resonance with a
positive $s$-wave scattering length \cite{exp,castin}.

\section{Ladder Resummation at Finite Density}

We now turn to the many-body problem of the repulsive Fermi gases.
The main purpose of this paper is to present a nonperturbative
calculation for function $f(x)$ from which we can give a better
prediction for the FMPT. In general, we expect that the new
nonperturbative result for $f(x)$ satisfies the following two
criteria: (i) The function $f(x)$ recovers the perturbative result,
Eq. (\ref{second}), at weak coupling $k_{\text F}a\rightarrow 0$;
and (ii) since we consider a short-range potential with $r_0\ll a$,
the physical result should be universal, i.e., $f(x)$ depends only
on the gas parameter $k_{\text F}a$ and does not depend on other
parameters such as the renormalization scale $\mu$. The criterion
(ii) is hard to fulfill since the loop corrections in quantum field
theory generally bring the renormalization scale dependence and
weaken the prediction power due to the arbitrariness in the choice
of the renormalization scale. However, in the following we show that
the result from the particle-particle ladder resummation,
corresponding to the leading order of the large-dimension expansion,
is independent of the arbitrary mass scale $\mu$.

\subsection{Nonperturbative energy density}
We first construct the nonperturbative version of the energy density
${\cal E}(x)$ using the vertex $C_0(\mu)$ determined in Sec. III and
the free propagators for the two spin components at finite density
\cite{fetter}
\begin{eqnarray}
{\cal G}_\sigma(p_0,{\bf p})=\frac{\Theta(|{\bf p}|-k_{\text
F}^\sigma)}{p_0-\omega_{\bf p}+i\epsilon}+\frac{\Theta(k_{\text
F}^\sigma-|{\bf p}|)}{p_0-\omega_{\bf p}-i\epsilon}.
\label{propagator}
\end{eqnarray}
Here $\sigma=\uparrow,\downarrow$, $k_{\text
F}^{\uparrow,\downarrow}=k_{\text F}\eta_{\uparrow,\downarrow}$ are
the Fermi momenta of the two spin components and $\Theta(z)$ is the
Heaviside step function. For each spin component, the propagator
(\ref{propagator}) describes two types of excitations, particles
with momentum $|{\bf p}|>k_{\text F}^\sigma$ and holes with $|{\bf
p}|<k_{\text F}^\sigma$.

The dilute imperfect Fermi gases are best described by resumming the
multiple interactions in terms of the scattering amplitude. The
Galitskii integral equations \cite{fetter} for the effective
two-particle interaction or scattering amplitude in the medium are
given by the ladder resummation \cite{Heisel}. On the other hand,
for large gas parameter $k_{\rm F}a$, one may look for other
expansion parameters instead of $k_{\rm F}a$ itself. Steele
\cite{resum1} and Sch{\"a}fer \emph{et al.} \cite{resum2} have
suggested a new expansion method using $1/{\cal D}$ as the expansion
parameter, where ${\cal D}=2^{D/2}$ with $D$ being the space-time
dimension. Most important, they have shown that the contribution of
the particle-particle (pp) ladder resummation, ${\cal
E}_{\text{int}}^{(0)}$ , is the leading-order contribution of the
$1/{\cal D}$ expansion \cite{resum1,resum2}, that is,
\begin{eqnarray}
{\cal E}={\cal E}_{\text{kin}}+{\cal E}_{\text{int}}^{(0)}+O(1/{\cal
D}).
\end{eqnarray}
All other contributions like hole-hole (hh) ladder sum and effective
range corrections are suppressed by a factor $1/{\cal D}$. According
to the above arguments, we expect that the most important
nonperturbative contributions come from the leading order of the
$1/{\cal D}$ expansion. The interaction energy density ${\cal
E}_{\text{int}}^{(0)}$ to this order is given by all the
particle-particle scattering terms (i.e., the $n$pp-$1$hh bubbles
for all $n=0,1,2,\cdots$).

To evaluate the interaction energy density, we first calculate the
elementary in-medium particle-particle bubble $B(P_0,{\bf P})$ shown
in Fig. \ref{fig1}(a). The fermion lines in the bubble diagram
correspond to the particle terms of the free propagator
(\ref{propagator}). According to the finite-density Feynmann rules
\cite{EFT}, it is given by
\begin{widetext}
\begin{eqnarray}
B(P_0,{\bf P})&=&i\int\frac{d^4q}{(2\pi)^4}\frac{\Theta(|{\bf
P}/2+{\bf q}|-k_{\text F}^\uparrow)}{\frac{P_0}{2}+q_0-\frac{({\bf
P}/2+{\bf q})^2}{2M}+i\epsilon}\frac{\Theta(|{\bf P}/2-{\bf
q}|-k_{\text F}^\downarrow)}{\frac{P_0}{2}-q_0-\frac{({\bf P}/2-{\bf
q})^2}{2M}+i\epsilon}\nonumber\\
&=&\int\frac{d^3{\bf q}}{(2\pi)^3}\frac{\Theta(|{\bf P}/2+{\bf
q}|-k_{\text F}^\uparrow)\Theta(|{\bf P}/2-{\bf q}|-k_{\text
F}^\downarrow)}{P_0-\frac{{\bf P}^2}{4M}-\frac{{\bf
q}^2}{M}+i\epsilon}\label{pp}.
\end{eqnarray}
For vanishing densities, $k_{\rm F}^\sigma=0$, the in-medium
particle-particle bubble recovers the vacuum result $B_0$. If the
on-shell condition is imposed, the in-medium particle-particle
bubble $B$ depends on not only the relative momentum ${\bf p}$ but
also the total momentum ${\bf P}$. This is due to the loss of
translational invariance in the presence of Fermi sea.

We can separate $B$ into a vacuum part and a medium part using the
identity
\begin{eqnarray}
\Theta(|{\bf P}/2+{\bf q}|-k_{\text F}^\uparrow)\Theta(|{\bf
P}/2-{\bf q}|-k_{\text F}^\downarrow) =1-\Theta(k_{\text
F}^\uparrow-|{\bf P}/2+{\bf q}|)-\Theta(k_{\text F}^\downarrow-|{\bf
P}/2-{\bf q}|) +\Theta(k_{\text F}^\uparrow-|{\bf P}/2+{\bf
q}|)\Theta(k_{\text F}^\downarrow-|{\bf P}/2-{\bf q}|).
\end{eqnarray}
The vacuum part (corresponding to $1$) is identical to $B_0$ defined
in the last section and is linearly divergent. The medium part is
convergent. For the vacuum part, it is natural to use the
dimensional regularization with PDS scheme introduced in the last
section.

Then the $n$pp-$1$hh bubble [see Fig. \ref{fig1}(b) for a typical
example] at given $n$ reads
\begin{eqnarray}
{\cal
E}_n&=&-C_0^{n+1}\int\frac{d^4P}{(2\pi)^4}\int\frac{d^4k}{(2\pi)^4}e^{i\eta
P_0}\frac{\Theta(k_{\text F}^\uparrow-|{\bf P}/2+{\bf
k}|)}{\frac{P_0}{2}+k_0-\frac{({\bf P}/2+{\bf
k})^2}{2M}-i\epsilon}\frac{\Theta(k_{\text F}^\downarrow-|{\bf
P}/2-{\bf k}|)}{\frac{P_0}{2}-k_0-\frac{({\bf P}/2-{\bf
k})^2}{2M}-i\epsilon}\left[B(P_0,{\bf
P})\right]^n\nonumber\\
&=&C_0^{n+1}\int\frac{d^3{\bf P}}{(2\pi)^3}\int\frac{d^3{\bf
k}}{(2\pi)^3}\Theta(k_{\text F}^\uparrow-|{\bf P}/2+{\bf
k}|)\Theta(k_{\text F}^\downarrow-|{\bf P}/2-{\bf
k}|)\int\frac{dP_0}{2\pi i}e^{i\eta P_0}\frac{\left[B(P_0,{\bf
P})\right]^n}{P_0-\frac{{\bf P}^2}{4M}-\frac{{\bf
k}^2}{M}-i\epsilon},
\end{eqnarray}
\end{widetext}
where $e^{i\eta P_0}$ with $\eta\rightarrow 0^+$ is a convergence
factor \cite{EFT}. The integration over $P_0$ picks up the pole or
imposes the on-shell condition $P_0={\bf P}^2/(4M)+{\bf k}^2/M$,
that is,
\begin{eqnarray}
\int\frac{dP_0}{2\pi i}e^{i\eta P_0}\frac{\left[B(P_0,{\bf
P})\right]^n}{P_0-\frac{{\bf P}^2}{4M}-\frac{{\bf
k}^2}{M}-i\epsilon}= \left[B({\bf P},{\bf k})\right]^n,
\end{eqnarray}
where the on-shell version of $B$ is given by
\begin{eqnarray}
&&B({\bf P},{\bf k})=\nonumber\\
&&M\int\frac{d^3{\bf q}}{(2\pi)^3}\frac{\Theta(|{\bf P}/2+{\bf
q}|-k_{\text F}^\uparrow)\Theta(|{\bf P}/2-{\bf q}|-k_{\text
F}^\downarrow)}{{\bf k}^2-{\bf q}^2+i\epsilon}.
\end{eqnarray}

The total interaction energy density ${\cal E}_{\text{int}}^{(0)}$
is given by
\begin{eqnarray}
{\cal E}_{\text{int}}^{(0)}=\sum_{n=0}^\infty{\cal E}_n.
\end{eqnarray}
Completing the summation of this geometric series, we obtain the
interaction energy density at the leading order of the $1/{\cal D}$
expansion,
\begin{eqnarray}
{\cal E}_{\text{int}}^{(0)}=C_0\int\frac{d^3{\bf
p}_1}{(2\pi)^3}\int\frac{d^3{\bf
p}_2}{(2\pi)^3}\frac{\Theta(k_{\text F}^\uparrow-|{\bf
p}_1|)\Theta(k_{\text F}^\downarrow-|{\bf p}_2|)}{1-C_0B({\bf
P},{\bf k})},
\end{eqnarray}
where ${\bf p}_{1,2}={\bf P}/2\pm{\bf k}$ as defined in Sec. III.
The imaginary part of $B$ can be evaluated as
\begin{eqnarray}
\text{Im}B({\bf P},{\bf k})=-\frac{M|{\bf k}|}{4\pi}\Theta(|{\bf
p}_1|-k_{\text F}^\uparrow)\Theta(|{\bf p}_2|-k_{\text
F}^\downarrow).
\end{eqnarray}
This quantity is nonzero only when the momenta ${\bf p}_1$ and ${\bf
p}_2$ are both above the Fermi surfaces. However, the final
integration over ${\bf p}_1$ and ${\bf p}_2$ in the interacting
energy density ${\cal E}_{\text{int}}$ is associated with a
phase-space factor $\Theta(k_{\text F}^\uparrow-|{\bf
p}_1|)\Theta(k_{\text F}^\downarrow-|{\bf p}_2|)$. Therefore, the
interaction energy density is real and physical, as we expected.

\begin{figure}[!htb]
\begin{center}
\includegraphics[width=9cm]{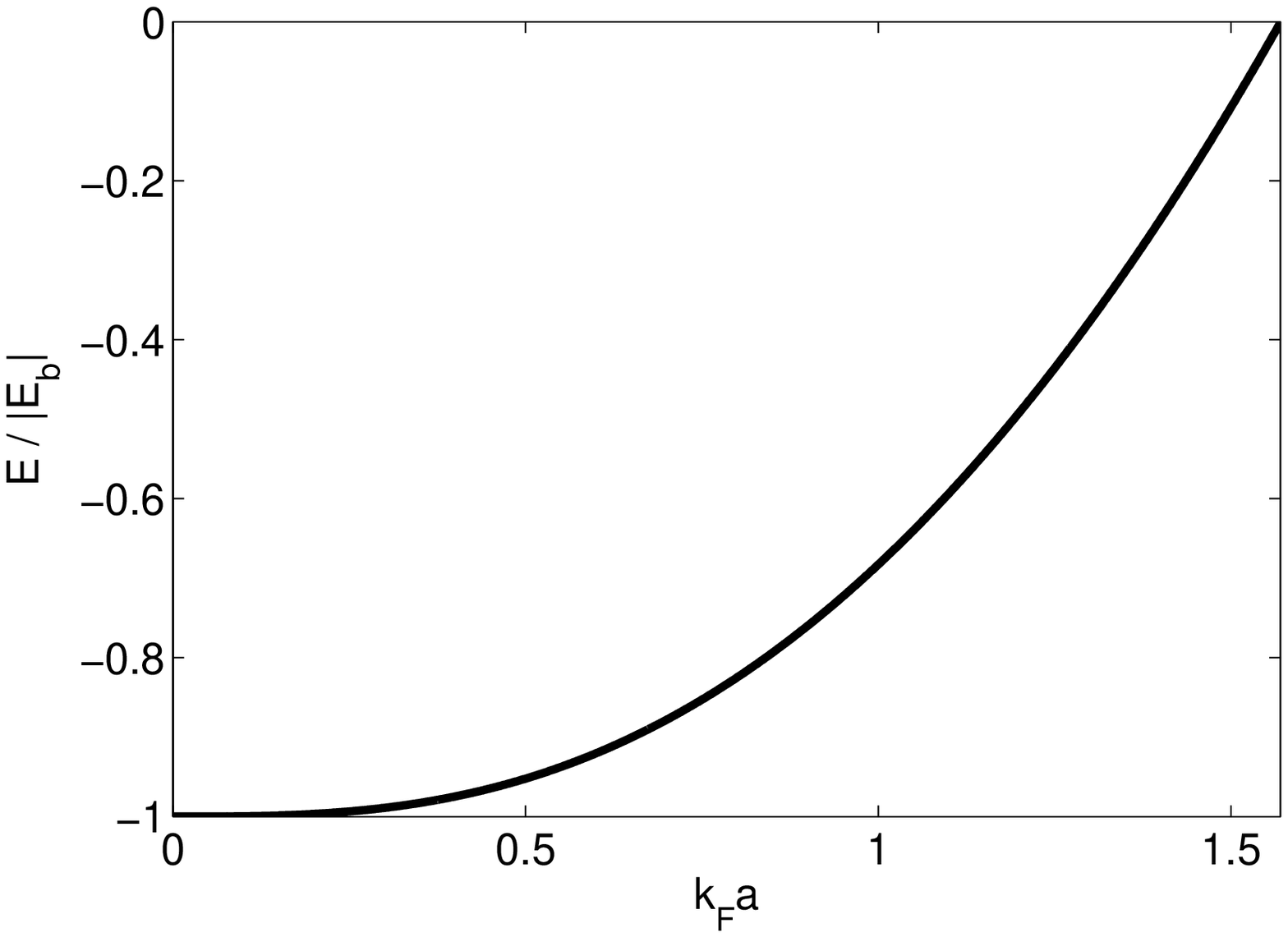}
\caption{The pole energy $E$ (divided by the binding energy $|E_{\rm
b}|=1/(Ma^2)$ in the vacuum) at zero pair momentum ${\bf P}=0$ as a
function of the gas parameter $k_{\rm F}a$. The pole energy turns
out to be positive for $k_{\rm F}a>\pi/2$.\label{figpole}}
\end{center}
\end{figure}

\begin{figure}[!htb]
\begin{center}
\includegraphics[width=9cm]{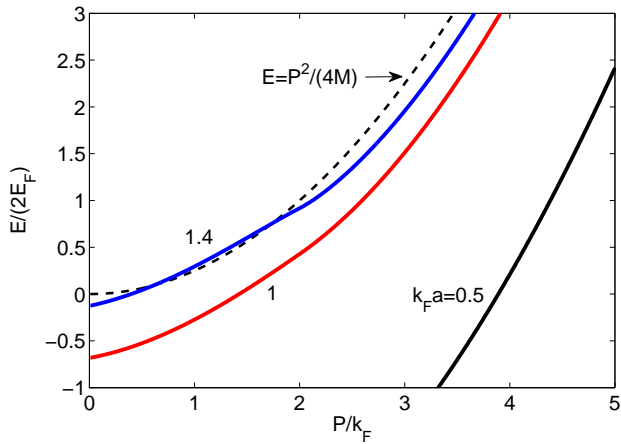}
\caption{(Color online) The pole energy $E$ (divided by $2E_{\rm
F}$) as a function of the pair momentum $P=|{\bf P}|$ (divided by
$k_{\rm F}$) for various values of the gas parameter $k_{\rm F}a$.
The dashed line corresponds to the dispersion
$E(P)=P^2/(4M)$.\label{figpole2}}
\end{center}
\end{figure}

\subsection{In-medium two-body problem}
Since we adopt a zero-range potential, a bound state with binding
energy $E_{\rm b}=-1/(Ma^2)$ always exists for positive scattering
length $a>0$. A key problem here is that how we can describe a
metastable repulsive Fermi gas where all fermions are forced on the
scattering states. Actually, the so-called upper branch has clear
meaning only in the two-body picture, and so far it is not clear to
what extent this two-body picture of a ``repulsive" Fermi gas will
persist. A recent study of three attractive fermions shows that
there are many nontrivial avoided crossings between the two branches
close to the resonance ($a\rightarrow\infty$), making it difficult
to unambiguously identify a repulsive Fermi system \cite{FMPT5}.

To realize a metastable repulsive Fermi gas we have to exclude the
molecule bound states of two atoms with unlike spins and enforce all
atoms to the scattering states \cite{Ho}. One possible prescription
is to subtract the contribution from the bound-state poles within
the Nozi$\acute{\rm e}$res-Schmitt-Rink (NSR) theory \cite{Ho}.
However, as designed, NSR theory works well only at temperature
higher than the critical temperature of superfluidity. The
particle-particle resummation theory we present complements the NSR
theory and can be regarded as the zero-temperature analog of the NSR
theory.

The key point in this problem is to consider the medium effects on
the bound-state properties. To this end, we first construct the pair
propagator ${\cal S}(P_0,{\bf P})$ in the presence of Fermi seas.
With the in-medium elementary particle-particle bubble $B(P_0,{\bf
P})$, the in-medium pair propagator ${\cal S}(P_0,{\bf P})$ is given
by the ladder resummation,
\begin{eqnarray}
{\cal S}(P_0,{\bf P})=\frac{C_0(\mu)}{1-C_0(\mu)B(P_0,{\bf P})}.
\end{eqnarray}
With this pair propagator, the interaction energy density ${\cal
E}_{\text{int}}^{(0)}$ can be expressed as
\begin{eqnarray}
{\cal E}_{\text{int}}^{(0)}&=&\int\frac{d^3{\bf
P}}{(2\pi)^3}\int\frac{d^3{\bf k}}{(2\pi)^3}\Theta(k_{\text
F}^\uparrow-|{\bf p}_1|)\Theta(k_{\text F}^\downarrow-|{\bf
p}_2|)\nonumber\\
&\times&\int\frac{dP_0}{2\pi i}e^{i\eta P_0}\frac{{\cal S}(P_0,{\bf
P})}{P_0-\frac{{\bf P}^2}{4M}-\frac{{\bf k}^2}{M}-i\epsilon}.
\end{eqnarray}

In general,  the in-medium pair propagator ${\cal S}(P_0,{\bf P})$
has a real pole $P_0=E({\bf P})$ corresponding to the in-medium
bound state. However, we now show that such pole does not contribute
to the energy density for the regime of the gas parameter $k_{\rm
F}a$ we are interested in. As shown in the last subsection, in the
calculation of the interaction energy density ${\cal
E}_{\text{int}}$, the on-shell condition $P_0={\bf P}^2/(4M)+{\bf
k}^2/M$ is imposed and the integrations over the momenta ${\bf P}$
and ${\bf k}$ are performed according to the finite-density Feynmann
rules. Therefore, if the energy dispersion of the pole $E({\bf P})$
satisfies the condition
\begin{eqnarray}
E({\bf P})<\frac{{\bf P^2}}{4M}
\end{eqnarray}
for arbitrary ${\bf P}$, its contribution to the energy density is
naturally excluded.

Since the main purpose of this paper is to study the FMPT which
corresponds to an instability toward a small polarization $x$, we
can set $x = 0$ here.  For convenience, we define two dimensionless
quantities $s=|{\bf P}|/(2k_{\rm F})$ and $z=\sqrt{MP_0-{\bf
P}^2/4+i\epsilon}/k_{\rm F}=\sqrt{P_0/(2E_{\rm F})-s^2+i\epsilon}$.
The in-medium pair propagator can be evaluated as
\begin{eqnarray}
{\cal S}(P_0,{\bf P})=\frac{4\pi}{M}\frac{1}{1/a-(k_{\rm
F}/\pi)W(s,z)},
\end{eqnarray}
where $W(s,z)$ is given by
\begin{widetext}
\begin{eqnarray}
W(s,z)&=&\left[1+s+z\ln\frac{1+s-z}{1+s+z}+\frac{1-s^2-z^2}{2s}\ln\frac{(1+s)^2-z^2}{1-s^2-z^2}\right]\Theta(1-s)\nonumber\\
&+&\left[2+z\ln\frac{(1-z)^2-s^2}{s^2-(1+z)^2}+\frac{1-s^2-z^2}{2s}\ln\frac{(1+s)^2-z^2}{(1-s)^2-z^2}\right]\Theta(s-1).
\end{eqnarray}
\end{widetext}
The same result was also obtained in a recent paper \cite{niemann}.

For zero pair momentum ${\bf P}=0$, the condition $E({\bf P})<{\bf
P}^2/(4M)$ implies $E(0)<0$.  We thus focus on the regime of the gas
parameter where the pole $E(0)$ is negative. In this case, $E(0)$ is
determined by a simple equation,
\begin{eqnarray}
\label{poleequation}
\frac{\pi}{2k_{\rm F}a}=1+\sqrt{\frac{-E(0)}{2E_{\rm
F}}}\arctan\sqrt{\frac{-E(0)}{2E_{\rm F}}}.
\end{eqnarray}
This equation has negative solution only for $0<k_{\text F}a<\pi/2$,
where the solution represents the binding energy of a in-medium
bound state. The numerical result for $E(0)$ is shown in Fig.
\ref{figpole}. In the low-density limit $k_{\rm F}\rightarrow0$,
$E(0)$ recovers the vacuum result $E_{\rm vac}(0)=E_{\rm
b}=-1/(Ma^2)$. However, at finite $k_{\rm F}$, the medium shields
the bound state and reduces the binding energy, that is,
$|E(0)|<|E_{\rm vac}(0)|$. For $k_{\rm F}a>\pi/2$ and $k_{\text
F}a<0$, Eq. (\ref{poleequation}) has a positive solution which
corresponds to the positive energy pole of the in-medium pair
propagator. Such a pole is associated with Cooper pairs, and its
appearance represents the BCS instability. This positive energy pole
does not lead to singularities in the energy density integration, as
can be seen in next subsection, and does not need special treatment.

Note that $E(0)<0$ is not a sufficient condition for $E({\bf
P})<{\bf P}^2/(4M)$. We thus have to check the energy dispersion
$E({\bf P})$ carefully. The numerical results for some values of the
gas parameter $k_{\rm F}a$ are shown in Fig. \ref{figpole2}. For
$k_{\rm F}a<1.34$, the condition $E({\bf P})<{\bf P}^2/(4M)$ is
fulfilled for all values of ${\bf P}$.  However, for $1.34<k_{\rm
F}a<\pi/2$, there exists a regime $P_1<|{\bf P}|<P_2$ where $E({\bf
P})>{\bf P}^2/(4M)$.

In conclusion, the condition $E({\bf P})<{\bf P}^2/(4M)$ is
fulfilled for $k_{\rm F}a<1.34$. Therefore, in the parameter regime
$k_{\rm F}a<1$ investigated in the following, the contribution from
the bound state can be naturally excluded in the ladder resummation
scheme.

\subsection{Evaluating the energy density}

Now we evaluate the explicit form of the energy density ${\cal
E}(x)$ and the dimensionless function $f(x)$. First, the elementary
particle-particle bubble $B({\bf P},{\bf k})$ can be decomposed into
four parts
\begin{eqnarray}
B({\bf P},{\bf k})=B_0({\bf P},{\bf k})+B_\uparrow({\bf P},{\bf
k})+B_\downarrow({\bf P},{\bf k})+B_{\uparrow\downarrow}({\bf
P},{\bf k}),
\end{eqnarray}
where $B_0$ is the vacuum part discussed in Sec. III and the other
parts are given by
\begin{widetext}
\begin{eqnarray}
&&B_\uparrow({\bf P},{\bf k})=-M\int\frac{d^3{\bf
q}}{(2\pi)^3}\frac{\Theta(k_{\text F}^\uparrow-|{\bf P}/2+{\bf
q}|)}{{\bf k}^2-{\bf q}^2+i\epsilon},\ \ \ \ B_\downarrow({\bf
P},{\bf k})=-M\int\frac{d^3{\bf q}}{(2\pi)^3}\frac{\Theta(k_{\text
F}^\uparrow-|{\bf P}/2-{\bf q}|)}{{\bf k}^2-{\bf q}^2+i\epsilon},\nonumber\\
&&B_{\uparrow\downarrow}({\bf P},{\bf k})= M\int\frac{d^3{\bf
q}}{(2\pi)^3}\frac{\Theta(k_{\text F}^\uparrow-|{\bf P}/2+{\bf
q}|)\Theta(k_{\text F}^\downarrow-|{\bf P}/2-{\bf q}|)}{{\bf
k}^2-{\bf q}^2+i\epsilon}.
\end{eqnarray}

For convenience, we define another dimensionless quantity $t=|{\bf
k}|/k_{\text F}$ together with $s$ as defined in the last
subsection. Since the imaginary part of $B$ does not contribute to
the interaction energy, we need only to evaluate the real part of
$B$. We have
\begin{eqnarray}
\text{Re}B_0(s,t)= -\frac{M\mu}{4\pi},\ \ \ \
\text{Re}B_\uparrow(s,t)=\frac{Mk_{\text
F}}{4\pi^2}R_\uparrow(s,t),\ \ \ \
\text{Re}B_\downarrow(s,t)=\frac{Mk_{\text
F}}{4\pi^2}R_\downarrow(s,t),\ \ \ \
\text{Re}B_{\uparrow\downarrow}(s,t)=\frac{Mk_{\text
F}}{4\pi^2}R_{\uparrow\downarrow}(s,t).
\end{eqnarray}
where $R_\sigma(s,t)$ ($\sigma=\uparrow,\downarrow$) reads
\begin{eqnarray}
R_\sigma(s,t)=\frac{\eta_\sigma^2-(s+t)^2}{4s}
\text{ln}\bigg|\frac{\eta_\sigma+s+t}{\eta_\sigma-s-t}\bigg|+\frac{\eta_\sigma^2-(s-t)^2}{4s}
\text{ln}\bigg|\frac{\eta_\sigma+s-t}{\eta_\sigma-s+t}\bigg|+\eta_\sigma,
\end{eqnarray}
and the function $R_{\uparrow\downarrow}(s,t)$ is
\begin{eqnarray}
R_{\uparrow\downarrow}(s,t)=\left\{ \begin{array} {r@{\quad,\quad}l}
 -\Theta(x)R_\downarrow(s,t)-\Theta(-x)R_\uparrow(s,t)&
 0<s<\frac{1}{2} |\eta_\uparrow-\eta_\downarrow|\\
 K_\uparrow(s,t)+ K_\downarrow(s,t)& \frac{1}{2}|\eta_\uparrow-\eta_\downarrow|<s<\frac{1}{2}|\eta_\uparrow+\eta_\downarrow| \\
 0 & \text{elsewhere.}
\end{array}
\right.
\end{eqnarray}
Here $K_\sigma(s,t)$ is defined as
\begin{eqnarray}
K_\sigma(s,t)=\frac{\eta_\sigma^2-s^2-t^2}{4s}\text{ln}\bigg|\frac{(\eta_\sigma-s)^2-t^2}{r^2-s^2-t^2}\bigg|+\frac{t}{2}\text{ln}\bigg|\frac{\eta_\sigma-s+t}{\eta_\sigma-s-t}\bigg|+\frac{s-\eta_\sigma}{2},
\end{eqnarray}
\end{widetext}
where $r^2=(\eta_\uparrow^2+\eta_\downarrow^2)/2$.

Finally, the elementary particle-particle bubble reads
\begin{eqnarray}
B(s,t)=-\frac{M\mu}{4\pi}+\frac{Mk_{\text
F}}{4\pi^2}R_{\text{pp}}(s,t),
\end{eqnarray}
where the function $R_{\text{pp}}(s,t)$ is defined as
\begin{eqnarray}
R_{\text{pp}}(s,t)=R_\uparrow(s,t)+R_\downarrow(s,t)+R_{\uparrow\downarrow}(s,t),
\end{eqnarray}
Substituting the result of $B(s,t)$ into the expression of ${\cal
E}_{\text{int}}^{(0)}$, we observe that the energy density is
independent of the renormalization mass scale $\mu$. Converting the
integration variables ${\bf p}_1$ and ${\bf p}_2$ to ${\bf P}$ and
${\bf k}$, we find that the function $f(x)$ can be expressed as
\begin{eqnarray}\label{fx}
f(x)=\frac{1}{2}(\eta_\uparrow^5+\eta_\downarrow^5)+\frac{80}{\pi}\int_0^\infty
s^2ds\int_0^\infty t dt I(s,t)F(s,t),
\end{eqnarray}
where $F(s,t)$ is given by
\begin{eqnarray}
F(s,t)=\frac{k_{\text F}a}{1-\frac{1}{\pi}k_{\text
F}aR_{\text{pp}}(s,t)}.
\end{eqnarray}
The function $I(s,t)$ appears due to integration over the angle
between ${\bf P}$ and ${\bf k}$. Its explicit form is
\begin{eqnarray}\label{Ifun}
I(s,t)=\Bigg[\frac{\eta_\uparrow^2-(s+t)^2}{4s}\Theta(s+t-\eta_\uparrow)
+(\eta_\uparrow\rightarrow\eta_\downarrow)+t\Bigg]\nonumber\\
\times\
\Theta(r^2-s^2-t^2)\Theta(\eta_\uparrow-|s-t|)\Theta(\eta_\downarrow-|s-t|).
\end{eqnarray}

As we mentioned in the beginning of this section, it is important to
check whether the present result for $f(x)$ is consistent with the
perturbative expression (\ref{second}) for weak coupling $k_{\rm
F}a\ll 1$. To this end, we expand the function $F(s,\kappa)$ as
\begin{eqnarray}
F(s,t)=k_{\text F}a+\frac{1}{\pi}(k_{\text
F}a)^2R_{\text{pp}}(s,t)+O((k_{\text F}a)^3).
\end{eqnarray}
Using the expressions for $I(s,t)$ and $R_{\text{pp}}(s,t)$, we can
show that
\begin{eqnarray}
\frac{80}{\pi}\int_0^\infty s^2ds\int_0^\infty t dt
I(s,t)=\frac{10}{9\pi}\eta_\uparrow^3\eta_\downarrow^3
\end{eqnarray}
and
\begin{eqnarray}
\frac{80}{\pi^2}\int_0^\infty s^2ds\int_0^\infty t dt
I(s,t)R_{\text{pp}}(s,t)=\frac{\xi(\eta_\uparrow,\eta_\downarrow)}{21\pi^2}.
\end{eqnarray}
Therefore, our nonperturbative expression (27) exactly recovers the
perturbative result (3) at weak coupling. This convinces us that the
present theoretical approach is suitable to study the universal
upper-branch Fermi gas with a positive scattering length. In
addition, we can compare the results from our theory and the
second-order perturbation on the same footing and study the
nonperturbative effects on the FMPT.

\subsection{Results and discussion}

\begin{figure}[!htb]
\begin{center}
\includegraphics[width=9cm]{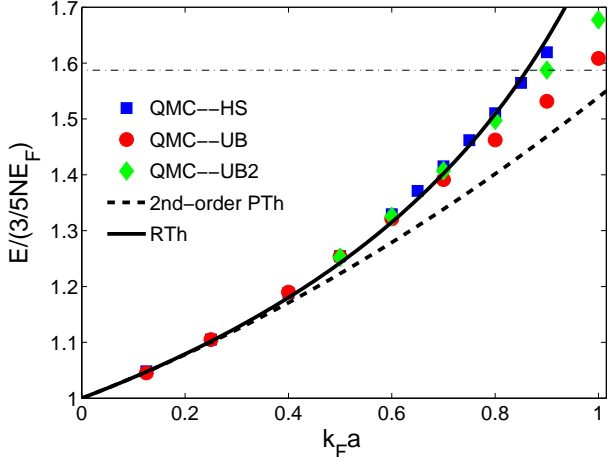}
\caption{(Color online) The energy density ${\cal E}$ (divided by
its value ${\cal E}_0=\frac{3}{5}nE_{\rm F}$ for noninteracting
Fermi gases) as a function of the gas parameter $k_{\rm F}a$
($0<k_{\rm F}a<1$) for the unpolarized case $x=0$. The solid line is
the result calculated from our particle-particle ladder resummation
theory (RTh). The dashed line is result of the second-order
perturbation theory (PTh). The dash-dotted horizontal line
corresponds to the energy of the fully polarized state ($x=1$),
i.e., $f(1)=2^{2/3}$. The blue squares are the quantum Monte Carlo
(QMC) data for the hard sphere (HS) potential \cite{QMC}, the red
circles are for the upper branch (UB) of a square well potential
\cite{QMC}, and the green diamonds are for the upper branch (UB2) of
an attractive short range potential \cite{QMC2}. For UB and UB2
cases, the effective range $r_0$ is much smaller than the $s$-wave
scattering length $a$ \cite{QMC,QMC2}. \label{fig2}}
\end{center}
\end{figure}

(A)\emph{Energy density and compressibility.} We first study the
equation of state for the unpolarized case $x=0$. The gas parameter
dependence of the energy density ${\cal E}$ in the regime $0<k_{\rm
F}a<1$ is shown in Fig. \ref{fig2}. We find that the result from the
ladder resummation is consistent with the perturbative result (3)
for small gas parameters $k_{\rm F}a<0.4$. However, significant
deviations are found for $k_{\rm F}a>0.4$, consistent with recent
quantum Monte Carlo simulations \cite{QMC,QMC2}. For the quantum
Monte Carlo simulations of the attractive interactions with a
negligible effective range (corresponding to UB and UB2 in Fig.
\ref{fig2}), the exclusion of molecular bound states is implemented
by choosing a two-body Jastrow factor \cite{QMC,QMC2} to be the
scattering solution of the attractive potential corresponding to
positive energy,  which, by construction, is orthogonal to the bound
molecules. Therefore, the accuracy of the quantum Monte Carlo data
depends on the choice of the Jastrow factor. Actually, exact
orthogonality of the many-body variational wave function to the
superfluid ground state (molecular condensation) can not be achieved
in the quantum Monte Carlo simulations \cite{QMC,QMC2}. We note that
our theoretical curve agrees better with the UB2 data than with the
UB data. The reason could be that the UB2 data from \cite{QMC2} are
obtained with a Jastrow factor which imposes a better orthogonality
to the superfluid ground state.

\begin{figure}[!htb]
\begin{center}
\includegraphics[width=9cm]{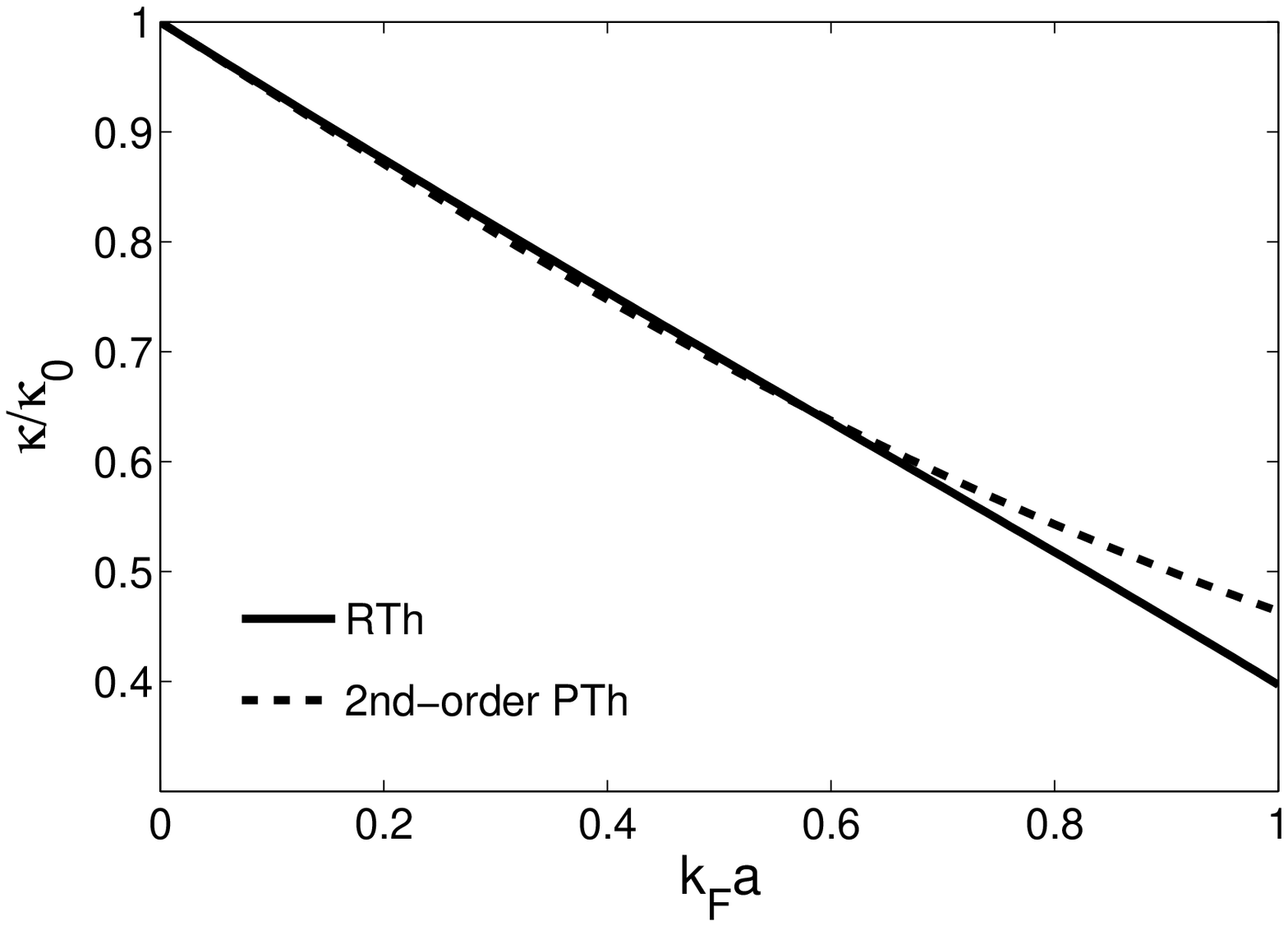}
\caption{The compressibility $\kappa$ [divided by its value
$\kappa_0=3/(2nE_{\rm F}]$ for non-interacting Fermi gases) as a
function of the gas parameter $k_{\rm F}a$ ($0<k_{\rm F}a<1$) for
the unpolarized case $x=0$. The solid line is the result calculated
from our particle-particle ladder resummation theory. The dashed
line is result of the second-order perturbation theory.
\label{fig3}}
\end{center}
\end{figure}

An important issue is whether the system is mechanically stable. The
mechanical stability of the system requires a positive
compressibility $\kappa$, which is defined as
\begin{eqnarray}
\frac{1}{\kappa}=n^2\frac{\partial^2{\cal E}}{\partial n^2}.
\end{eqnarray}
For the present ladder resummation theory, the explicit form of
$\kappa$ can be evaluated as
\begin{eqnarray}
\frac{\kappa_0}{\kappa}=1+\frac{144}{\pi}\int_0^\infty
s^2ds\int_0^\infty t dt I(s,t) G(s,t),
\end{eqnarray}
where $\kappa_0=3/(2nE_{\rm F})$ is the compressibility for
noninteracting Fermi gases, and the function $G(s,t)$ is given by
\begin{eqnarray}
G(s,t)=F(s,t)+\frac{5R_{\text{pp}}(s,t)}{9\pi}F^2(s,t)+\frac{R_{\text{pp}}^2(s,t)}{9\pi^2}F^3(s,t).
\end{eqnarray}
The compressibility $\kappa$ as a function of the gas parameter
$k_{\rm F}a$ is shown in Fig. \ref{fig3}. Comparing to the result
from the second-order perturbation theory,
\begin{eqnarray}
\frac{\kappa_0}{\kappa}=1+\frac{2}{\pi}k_{\text
F}a+\frac{8(11-2\ln2)}{15\pi^2}(k_{\text F}a)^2,
\end{eqnarray}
good agreement is found for small gas parameters, as we expected. In
the regime $0<k_{\rm F}a<1$ we are interested in, we find that the
compressibility $\kappa$ is positive, indicating that the system is
mechanically stable.

(B)\emph{Spin susceptibility.} Next we study the response of the
energy density to an infinitesimal polarization $x$. This response
is referred to as the spin (or magnetic) susceptibility. The spin
susceptibility $\chi$ can be defined as
\begin{eqnarray}
\frac{1}{\chi}=\frac{1}{n^2}\frac{\partial^2{\cal E}}{\partial
x^2}\bigg|_{x=0}=\frac{3E_{\rm F}}{5n}\frac{\partial^2f(x)}{\partial
x^2}\bigg|_{x=0}.
\end{eqnarray}
In the present ladder resummation theory, an explicit form of $\chi$
is hard to obtain. In practice, we expand the function $f(x)$ near
$x=0$ as $f(x)=f(0)+\alpha x^2+\cdots$. The coefficient $\alpha$ is
related to the spin susceptibility by
\begin{eqnarray}
\frac{\chi_0}{\chi}=\frac{9}{5}\alpha,
\end{eqnarray}
where $\chi_0=3n/(2E_{\text F})$ is the spin susceptibility of
noninteracting Fermi gases. Therefore, a diverging spin
susceptibility generally indicates a FMPT, as long as the transition
is of second order.

In the second-order perturbation theory, an analytical result for
$\chi$ can be achieved,
\begin{eqnarray}
\frac{\chi_0}{\chi}=1-\frac{2}{\pi}k_{\text
F}a-\frac{16(2+\text{ln}2)}{15\pi^2}(k_{\text F}a)^2,
\end{eqnarray}
which indicates a diverging spin susceptibility at $k_{\text
F}a=1.058$. However, this differs from the critical gas parameter
$(k_{\text F}a)_c=1.054$,  because the phase transition is of first
order in the second-order perturbation theory due to the appearance
of the nonanalytical term $\upsilon x^4\ln|x|$ with $\upsilon>0$.

\begin{figure}[!htb]
\begin{center}
\includegraphics[width=9cm]{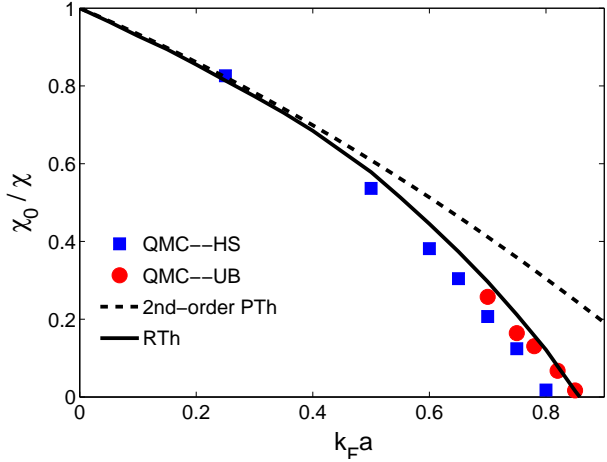}
\caption{(Color online) The dimensionless inverse spin
susceptibility $\chi_0/\chi$ as a function of the gas parameter
$k_{\rm F}a$. The blue squares and red circles are the QMC data
\cite{QMC} for the HS and UB cases, respectively. The solid line is
the result calculated from the particle-particle ladder resummation.
The dashed line is the second-order perturbative result.
 \label{fig4}}
\end{center}
\end{figure}

Our result for the spin susceptibility $\chi$ as a function of the
gas parameter $k_{\rm F}a$ is shown in Fig. \ref{fig4} and compared
with the perturbative result. We find that the spin susceptibility
predicted by the ladder resummation deviates significantly from the
second-order perturbative result for $k_{\text F}a>0.4$. Further,
the spin susceptibility diverges at $k_{\rm F}a=0.858$, in contrast
to the value $1.058$ from the second-order perturbation theory. The
data from the quantum Monte Carlo simulations \cite{QMC} are also
shown in Fig. \ref{fig4} as a comparison. Our theoretical result is
in good agreement with the data for the upper branch of the square
well potential where the effective range $r_0$ is tuned to be much
smaller than the scattering length \cite{QMC}. The gas parameter
$k_{\rm F}a=0.86$ where $\chi$ diverges is very close to our
prediction $k_{\rm F}a=0.858$. For the purely repulsive potential,
that is, the hard-sphere potential, the effective range effect
cannot be neglected \emph{a priori}. However, we find that our
result still has nice agreement with the data for the hard-sphere
case. The gas parameter $k_{\rm F}a=0.82$ where $\chi$ diverges is
also close to our prediction $k_{\rm F}a=0.858$. Actually, the
difference between the upper branch and the hard sphere cases [i.e.,
$0.86-0.82=0.04$] is very small compared with the critical gas
parameters. This indicates that the contribution from the effective
range effect is relatively small even for $k_{\rm F}a\sim O(1)$, if
the quantum Monte Carlo results are reliable. This can be understood
from the large-dimension expansion \cite{resum1,resum2} introduced
in the beginning of this section: The particle-particle ladder sum
is the leading-order contribution in the $1/{\cal D}$ expansion, and
all other contributions including the effective range corrections
are suppressed by a factor $1/{\cal D}$.

(C)\emph{Ferromagnetic transition.} While a diverging spin
susceptibility indicates a ferromagnetic phase transition, the order
of the ferromagnetic phase transition and the critical gas parameter
$(k_{\text F}a)_c$ should be obtained by studying carefully the
shape of the energy landscape, that is, the full $x$ dependence of
the function $f(x)$. To very high numerical accuracy, we have not
found any maximum at $x\neq0$ in the energy landscape. Instead, we
find a second-order phase transition at $k_{\text F}a=0.858$, where
the function $f(x)$ starts to develop a minimum at $x\neq0$,
consistent with the gas parameter where the spin susceptibility
diverges. This is in contrast to the second-order perturbation
theory which predicts a first-order phase transition at $k_{\text
F}a=1.054$ \cite{FMPT2}, where the spin polarization $x$ jumps from
zero to $x_c=0.573$. A second-order FMPT for a zero-range potential
model was also obtained by Heiselberg \cite{FMPT6} recently using a
completely different many-body method.

It seems that our result of a second-order phase transition is in
contradiction to the BKV argument \cite{Belitz}. However, the BKV
argument is based on the assumption that $\upsilon>0$. Actually, we
have fitted the energy density of the form $f(x)=f(0)+\alpha
x^2+\upsilon x^4\text{ln}|x|+\beta x^4$ for small $x$. For small gas
parameter $k_{\text F}a<0.3$, the coefficient $\upsilon$ agrees well
with the perturbative result $\upsilon=40(k_{\text
F}a)^2/(243\pi^2)$. However, for larger $k_{\text F}a$ (especially
around the critical gas parameter), it turns out to be negative due
to the nonperturbative effects. This indicates that the FMPT in the
systems of dilute repulsive Fermi gases corresponds to the case
$\upsilon<0$ and is a counterexample to the BKV argument where the
assumption $\upsilon>0$ is adopted.

Since an analytical expression for the function $f(x)$ as well as
the coefficient $\upsilon$ cannot be achieved in the present ladder
resummation theory, we cannot understand analytically how the
nonparturbative effects modify the order of the phase transition. In
fact, analytical results cannot be obtained from the order
$O((k_{\text F}a)^3)$ even for the unpolarized case $x=0$ in the
perturbation theory \cite{EFT}. However, some definite conclusions
can be drawn from our numerical results: (1) Higher-order terms in
the gas parameter can also generate nonanalytical terms of the form
$x^4\text{ln}|x|$ and may generate other important non-analytical
terms which are not known due to the mathematical limitation. (2)
The coefficients of the nonanalytical terms generated by the
higher-order contributions are certainly not always positive, and
they are generally proportional to $(k_{\text F}a)^n$ for the
$n$th-order contributions. Since the phase transition occurs at a
gas parameter $k_{\text F}a\sim O(1)$, the nonperturbative effects
from the sum of the higher order contributions are very important.
As we have shown numerically, their effects are not only reducing
the critical value of the gas parameter but also changing the order
of the phase transition.

\begin{figure}[!htb]
\begin{center}
\includegraphics[width=9cm]{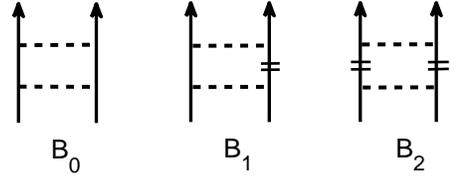}
\caption{The elementary bubbles organized in the number of the MI.
The solid line with a cut represents the MI part of the propagator,
and the pure solid line corresponds to the vacuum part.
\label{fig5}}
\end{center}
\end{figure}

\section{Inclusion of Hole-Hole Ladders}

In this section we check whether our conclusion that the FMPT is of
second order is changed by other contributions. We consider the
contributions from the hole-hole ladder diagrams by summing the
combined particle-particle and hole-hole ladders to all orders in
$k_{\text F}a$ while keeping the criteria (i) and (ii)  satisfied.
Such a resummation scheme for the unpolarized case $x=0$ has been
performed by Kaiser \cite{resum3}.

Following the treatment by Kaiser \cite{resum3}, we rewrite the
propagator (\ref{propagator}) in an alternative form
\begin{eqnarray}
{\cal G}_\sigma(p_0,{\bf p})={\cal G}_0(p_0,{\bf p})+2\pi
i\delta(p_0-\omega_{\bf p})\Theta(k_{\text F}^\sigma-|{\bf p}|),
\end{eqnarray}
where the first term corresponds to the vacuum propagator ${\cal
G}_0(p_0,{\bf p})=(p_0-\omega_{\bf p}+i\epsilon)^{-1}$ and the
second term is a so-called medium insertion (MI) \cite{resum3}. The
elementary bubbles in this treatment are shown in Fig. \ref{fig5}.
The first diagram $B_0$ is identical to the vacuum part studied in
Sec. III and it can be renormalized using the PDS scheme. For our
purpose of resummation, we are interested in the following two
quantities, $B_0+B_1+B_2$ and $B_0+B_1$, which are mutually complex
conjugate. We have
\begin{eqnarray}
B_0+B_1+B_2&=&-\frac{M\mu}{4\pi}+\frac{Mk_{\text
F}}{4\pi^2}\left[R(s,t)-i\pi I(s,t)\right],\nonumber\\
B_0+B_1&=&-\frac{M\mu}{4\pi}+\frac{Mk_{\text
F}}{4\pi^2}\left[R(s,t)+i\pi I(s,t)\right],
\end{eqnarray}
where $R(s,t)=R_\uparrow(s,t)+R_\downarrow(s,t)$ and $I(s,t)$ is the
function defined in (30).

To sum all ladder diagrams built from the elementary bubbles, we
first notice that the nonvanishing contributions to the interaction
energy come from diagrams with at least two adjacent MIs
\cite{resum3}. Then a typical $n$th-order contribution would look
like the ring diagram of Fig. \ref{fig1} (b) with $n$ vertices and
at least two adjacent MIs. Naively, all these $n$th-order diagrams
are summed to give $g^n[(B_0+B_1+B_2)^n-(B_0+B_1)^n]$, where the
subtraction gets rid of those diagrams which have no adjacent MI
pairs. However, this expression is complex and therefore cannot be
the correct one. The crucial observations are that as follows: (1)
Each $n$th-order ring diagram has an $n$-rotational symmetry.
Therefore, we should introduce an additional factor $1/n$. (2) An
$n$th-order ring diagram comes from closing two open MI lines of an
$n$th-order ladder diagram, which introduces an integration over the
allowed phase space $|{\bf p}_1|<k_{\text F}^\uparrow$ and $|{\bf
p}_2|<k_{\text F}^\downarrow$ but does not contribute a factor $B_2$
to the energy as the naive expression does. These amendments lead to
the correct $n$th-order contribution to the interaction energy
\cite{resum3}: $g^n[(B_0+B_1+B_2)^n-(B_0+B_1)^n]/(2iIn)$. The
summation over $n$ leads to two complex-conjugated logarithms and
the final result is real.

The final result for the energy density does not depend on the
renormalization scale $\mu$, and the function $f(x)$ in this
resummation scheme also takes the form (\ref{fx}), while the
function $F(s,t)$ becomes
\begin{eqnarray}
F(s,t)=\frac{\text{ln}\left[1-\frac{1}{\pi}k_{\text F}a
R(s,t)+ik_{\text F}a I(s,t)\right]-\rm{c.c.}}{2iI(s,t)}.
\end{eqnarray}
For small gas parameter $k_{\rm F}a\ll1$, $F(s,t)$ can be expanded
as
\begin{eqnarray}
F(s,t)=k_{\text F}a+\frac{1}{\pi}(k_{\text F}a)^2R(s,t)+O((k_{\text
F}a)^3).
\end{eqnarray}
We can also check that
\begin{eqnarray}
\frac{80}{\pi^2}\int_0^\infty s^2ds\int_0^\infty t dt
I(s,t)R(s,t)=\frac{\xi(\eta_\uparrow,\eta_\downarrow)}{21\pi^2},
\end{eqnarray}
which reflects the fact that the hole-hole ladders start to
contribute at the order $O((k_{\text F}a)^3)$
\cite{EFT,resum1,resum2}. Therefore, the criteria (i) and (ii) are
also fulfilled in the present resummation theory. Numerically, we
also find a second-order phase transition, which occurs at a smaller
gas parameter $k_{\text F}a=0.786$. We note that the inclusion of
hole-hole ladders may not improve the quantitative result, since it
only includes part of the beyond-leading-order contribution in the
large-${\cal D}$ expansion.

\section{Summary}

In summary, we have studied the nonperturbative effects on the
ferromagnetic phase transition in repulsive Fermi gases by summing
the ladder diagrams to all orders in the gas parameter $k_{\text
F}a$. The nonperturbative effects not only reduce the critical gas
parameter but also change the order of the phase transition. The
resummation of particle-particle ladders, which corresponds to the
leading order of the large-dimension expansion, predicts a
second-order phase transition occurring at $k_{\text F}a=0.858$, in
good agreement with the quantum Monte Carlo result \cite{QMC}. The
spin susceptibility calculated from our resummation theory is also
in good agreement with the quantum Monte Carlo results. Therefore,
the resummation of the ladder diagrams provides a more quantitative
way to study the ferromagnetic transition in repulsive Fermi gases.
In this paper, we have considered only a zero-range potential model.
It will be interesting to study the nonuniversal shape-dependent
contributions using the finite-density effective range expansion
\cite{resum2}.

{\bf Acknowledgments:}\ We thank S. Pilati and S.-Y. Chang for
providing us with the QMC data, N. Kaiser for helpful
communications, and A. Sedrakian for reading the manuscript. L. He
acknowledges the support from the Alexander von Humboldt Foundation,
and X.-G. Huang is supported by the Deutsche Forschungsgemeinschaft
(Grant SE 1836/1-2).

\end{document}